# The Importance of Being Honest: Correlating Self-Report Accuracy and Network Centrality with Academic Performance

Fronzetti Colladon, A., & Grippa, F.

# The Importance of Being Honest. Correlating Self-Report Accuracy and Network Centrality with Academic Performance

Fronzetti Colladon, A., & Grippa, F.

**Abstract**

This study investigates the correlation of self-report accuracy with academic performance. The sample was composed of 289 undergraduate students (96 senior and 193 junior) enrolled in two engineering classes. Age ranged between 22 and 24 years, with a slight over representation of male students (53%). Academic performance was calculated based on students' final grades in each class. The tendency to report inaccurate information was measured at the end of the Raven Progressive Matrices Test, by asking students to report their exact finishing times. We controlled for gender, age, personality traits, intelligence, and past academic performance. We also included measures of centrality in their friendship, advice and trust networks. Correlation and multiple regression analyses results indicate that lower achieving students were significantly less accurate in self-reporting data. We also found that being more central in the advice network was correlated with higher performance ($r = .20$, $p < .001$). The results are aligned with existing literature emphasizing the individual and relational factors associated with academic performance and, pending future studies, may be utilized to include a new metric of self-report accuracy that is not dependent on academic records.

## Introduction

For the past few decades social psychology researchers have focused their attention on the association between personality and academic accomplishments (Busato, Prins, Elshout, & Hamaker, 2000; Chamorro-Premuzic & Furnham, 2008; Lievens, Ones, & Dilchert, 2009; Poropat, 2014). Factors such as intelligence and conscientiousness have been thoroughly explored in connection to academic achievement. Less explored factors, such as self-report accuracy and the strengths of ties within social networks, have been attracting the attention of researchers within higher education literature (Gonyea, 2005; Rourke & Kanuka, 2009). More research is still necessary, in particular to explore the predictive value of accuracy and centrality on academic success. Most of the studies exploring the correlation between accuracy and performance have focused on the ability of individuals to accurately self-report their grades - i.e. Graduate Record Examination (GRE) and Grade Point Average (GPA) (Cole & Gonyea, 2010; Kirk & Sereda, 1969). While previous studies have measured the ability or tendency of students to misrepresent their grades (e.g. "What is your GPA"), in this study we explore how accurately students reported the finishing time on a test. Students reported the finishing time of the Raven Progressive Matrices Test (J. J. J. Raven, Raven, John, & Raven, 2003), which was immediately verified by the instructors on the test website. The goal of our study is to contribute to the literature on self-report accuracy by exploring the correlations between performance and accuracy at self-reporting "non-academically related data". Differently from the majority of previous studies, we use the accuracy in reporting the time on a cognitive test, instead of reporting their academic record. We have seen an increasing interest in studying the impact of measures of personality on academic performance, looking in particular at factors such as intelligence and the Big Five traits. In a meta-analysis of other-rated personality, Poropat (2014) found that personality variables are important correlates of academic performance, especially conscientiousness, emotional stability, and openness to experience, while GPA correlations with conscientiousness exceeded those with intelligence. Recent studies consistently showed positive correlation between conscientiousness and academic performance (Battistoni & Fronzetti Colladon, 2014; Conard, 2006; Lievens et al., 2009; Uppal, 2014), though the correlations with openness to experience, neuroticism, agreeableness and extraversion have been not always consistent and significant (Busato et al., 2000; Chamorro-Premuzic & Furnham, 2008; Conard, 2006). Other studies suggest that people perform better when they demonstrate personality traits such as self-monitoring or conscientiousness, which are highly respected in the work environment (Barrick, Mount, & Judge, 2001; Fang et al., 2015). For example, Chamorro-Premuzic and Furnham (2008) found that conscientiousness and fluid intelligence are predictors of academic performance. Fluid intelligence is the general ability to think abstractly, identify patterns, solve problems, and discern relationships and personality. Several studies found that intelligence is a predictor of both academic and organizational performance (Chamorro-Premuzic & Furnham, 2008; Gottfredson, 1997; Joseph, Jin, Newman, & Boyle, 2015; Judge, Bono, Ilies, & Gerhardt, 2002).

Although self-report accuracy could be partly explained looking at individual personality traits, in this study we did not focus on the associations between academic performance and personality traits, or between these traits and accuracy at reporting data. Our goal is to specifically explore the associations between being accurate in reporting an exam's finishing time and academic performance.

A second contribution of this study is to explore how academic performance is associated to individuals' network centrality. Network centrality is conceptualized as the degree of connection between individuals, which offers insights into their relative prominence and influence over others (Everett & Borgatti, 2005; Freeman, 1979). In this paper, we map ties representing requests for advices, number of friends, or number of classmates considered trustworthy. A high number of direct contacts can translate into a high number of incoming and outgoing network ties for a social actor. If two students are not directly connected (no incoming or outgoing ties), they can still rely on others who are more connected to the rest of the class in order to exchange information, emotional support or advice. As we describe in the method section, we operationalize individuals' ability to mediate indirect connections among peers using the metric of betweenness centrality and we use degree centrality to measure the number of connections to friends, to peers to rely on for advice seeking and number of ties to trustworthy peers (Freeman, 1979).

## Self-Report Accuracy

Self-report data is widely used by social scientists in surveys and questionnaires. It has been classified in two categories: attitudinal data (e.g. "How satisfied are you with your learning experience") and factual data (e.g. "What is your current GPA?"). Self-reported factual data can be verified against external data sources such as records held by the institution (Gonyea, 2005). For example, Cole and Gonyea (2010) found that lower achieving students are much less accurate when reporting their scores in important tests such as the Scholastic Assessment Test (SAT), and the American College Testing (ACT). The lack of precision was such that students' self-reports of SAT scores were consistently higher than actual scores by more than 20 points.

The use of self-reported data introduces problems of validity and reliability, due to cognitive and social desirability biases, that are well studied in literature (Leite & Beretvas, 2005; Porter, 2011; Tracey, 2016). The various explanations for error in accuracy have been associated to cognitive distortion or motivated distortion (Willard & Gramzow, 2008). The cognitive distortion approach is related to difficulties in encoding experiences in the long term memory or in retrieving memories. On the contrary, inaccuracy due to motivated distortion is described as an intentional report of incorrect data, often based on a social desirability bias that prompts individuals to change a response before communicating it, in order to project an inaccurate image of academic performance (Tracey, 2016).

In our study, we expect a social desirability bias rather than a cognitive one, since students were asked to report the finishing time of the test immediately after they had completed this task. A study from the Center for Academic Integrity in the US indicates that the main motivation for students to provide incorrect information about their performance is to raise their grades (D. L. McCabe, Treviño, & Butterfield, 2002). Anaya (1999) studied the accuracy of self-reported GRE scores against institutional records and found a strong correlation between self-reported and actual GRE scores ($r = .94$). The same study found that higher achieving students and female students were considerably more accurate when reporting their own scores. Mayer et al. (2007) found evidences of a systematic bias towards over-reporting, as students overestimated their actual SAT scores by an average of 25 points, with 10% under-reporting, 51% reporting accurately, and 39% over-reporting. In another study, Finn and Frone (2004) demonstrated that behaviors such as cheating are more likely to occur among lower achieving students, especially when they do not identify with the institution, and among higher achieving students with low levels of academic self-efficacy (i.e. evaluation of their own ability to perform a task). Similarly,

Rosen, Porter and Rogers (2017) found that lower-performing 9th grade students are less accurate than higher-performing students when they reported on the grade in their Algebra course.

Although previous studies had not measured accuracy on reporting a variable like "the finishing time", we would expect a similar positive association between accuracy and performance on the final exam. We expect students to be apprehensive during a cognitive test, which could make them inclined to provide deceptive information in order to look more competent. Students might assume that if they finish the test earlier, the instructor will think highly of them.

Asking students to self-report academic data (e.g. GPA, GRE, STA, ACT) is used extensively in higher education, although its validity has been debated for more than forty years (Pike, 2011). There are potential methodological problems with using metrics of self-report accuracy that are dependent on academic records. The primary concern with self-reported factual data is concurrent validity, which refers to the extent to which the results of a measurement correlates to the results of a previously established measurement for the same construct (Creswell, 2012). Instead of measuring accuracy by looking at self-reported academic records, we use the accuracy in reporting the finishing time on a cognitive test, which might not necessarily trigger a social desirability effect. By doing so, we correlate a new metric, accuracy in reporting a test finishing time, with academic performance. There are potential advantages for using this metric or similar self-report accuracy indicators that are not dependent on academic records. As suggested by a recent longitudinal study of more than 23,000 9th graders in U.S. secondary schools (Rosen, Porter, & Rogers, 2017), students are fairly good reporters of course-taking patterns but poor reporters of more potentially sensitive questions. Reporting a finishing time is not necessarily a sensitive question, and can be interpreted by students as a fairly unobtrusive request. Students should have no incentive in misrepresenting reality, by reporting a different time. Only students who rationally decide to project a different image should have a high discrepancy in the reported time. In order to explore how academic performance is associated with this new accuracy indicator, we propose the following hypothesis:

*H1: Students who are more accurate at self-reporting the finishing time on a test are also performing better academically.*

**Centrality in Social Networks**

Individuals' roles and behaviors in a social network can derive from several sources: trust, degree of acquaintance/friendship and advice seeking relationships (Buskens, 1998; Fronzetti Colladon et al., 2017). In this section we describe how being well connected within social networks that are based on friendship, advice-seeking and trust is associated to academic performance, either directly or through mediating variables like individual's roles, behaviors and traits. The advice network has been defined as a set of "*relations through which individuals share resources such as information, assistance, and guidance*" (Sparrowe, Liden, Wayne, & Kraimer, 2001). Being central in the advice network means receiving more requests for help and tapping into other people's knowledge. Individuals with a high in-degree centrality in the advice network are usually preferred for their input on a specific competence area, while actors who have a high in-degree centrality in the friendship network are chosen for their companionship (Klein, Lim, Saltz, & Mayer, 2004). The friendship network is defined by the ties of affection and companionship that connect individuals (Baldwin, Bedell, & Johnson, 1997). Being central in the friendship network means having access to emotional support and has been demonstrated as affecting student popularity (Fronzetti Colladon, Grippa, Battistoni, Gloor, & La Bella, 2017). The trust network identifies the relations that connect individuals based on reliability and confidence. Being central in the trust network implies that peers depend and rely on you and that you rely on others as well. Friendship and trust are closely related constructs, with friendship dependent on trust to grow, and trust often used as a conceptual base for friendship (Warris & Rafique, 2009). Friendships rest on intimacy and trust rather than on existing task structures and competences (Gibbons, 2004).

Fang et al. (2015) recently found that individuals' personality and their social network centrality are both positively associated with performance and career success. Similarly, Sparrowe, Liden, Wayne and Kraimer (2001) found a positive association between job performance and centrality in the advice networks. Battistoni and Fronzetti Colladon (2014) focused on the correlation between personality traits and social network position of college students with regard to the advice networks. Their study provided empirical evidence of significant associations between key network positions and academic performance and traits such as conscientiousness, neuroticism and agreeableness. Mehra, Kilduff and Brass (2001) demonstrated that being central in the advice network is associated with higher academic achievement. Similarly, Cattani and Ferriani (2008) provided evidence of the importance of holding a central position in informal social networks: central individuals are more likely to acquire and integrate new knowledge from multiple sources with clear positive impact on performance. Trust has been often cited as an essential and important element of a successful social network (Edwards & Grinter, 2001; Sherchan, Nepal, & Paris, 2013). Individuals who are sought after for advice are usually considered the experts, though not necessarily the most trustworthy (Krackhardt & Hanson, 1993). Other studies indicate that competence, affect-based trust and social interaction impact task-related interactions (Hinds, Carley, Krackhardt, & Wholey, 2000). Empirical evidences on the effect of interpersonal trust on academic performance seem to provide mixed results. For example, Goddard, Salloum and Berebitsky (2009) found that trust is a strong predictor of several important outcomes including student achievement, though they use a proxy variable as they refer to a general "level of trust towards the institution" which leads teachers to feel greater responsibility and invest in students' academic success.

Research on relatedness and academic performance has examined the effects of students' connectedness to particular social partners, particularly teachers, parents, and peers (Dunbar, Dingel, Dame, Winchip, & Petzold, 2016; Furrer & Skinner, 2003). Students who feel disconnected or rejected by key partners are more likely to experience frustration, lack of engagement and alienation from learning activities, which interferes with their academic performance (Wentzel, 2005). Other studies indicate that the perception of peer support is related to academic performance (Buhs, 2005) and academic motivation (Altermatt & Pomerantz, 2003; Furrer & Skinner, 2003). Peer support has been usually studied with reference to the number of friends, the quality of the relations (Berndt, 2002) or the type of support, either academic, emotional, or social (Juvonen, 2007). Students tend to connect with others exhibiting similar attributes in terms of specific attributes, usually age and gender. This phenomenon is known as network homophily (Mcpherson, Smith-Lovin, & Cook, 2001). In general, homophily represents the idea that individual's personal networks are homogeneous with regard to many sociodemographic, behavioral, and intrapersonal characteristics (Mcpherson et al., 2001).

Previous studies focused on the impact of ties within one specific social network, looking at the impact of friendship or advice (Mehra et al., 2001; Tasselli & Kilduff, 2017). With reference to the impact of friendship on academic performance, previous studies do not clearly support the conclusion that individual performance is higher for individuals who are central in the friendship network. While there is no clear evidence in literature that stronger friendship ties with peers or colleagues lead to greater academic success (Baldwin et al., 1997), we would expect that academic success depends on the ability to rely on peers for emotional support (trust, friendship) or for accessing important knowledge. In her recent book on the importance of friendship for success in school, McCabe (2016) suggested that having a good friend affects a child's school performance, since students with more friends tend to have better attitudes about school and learning than kids who report fewer friends. Having a lot of friends to rely on provides students with emotional support that could positively influence academic focus and self-esteem (Flashman, 2012).

Our second hypothesis considers separately the impact of student centrality in the advice, friendship, and trust networks. The main reason is that while some studies indicate a strong connection between performance and being central in the advice network (Cattani & Ferriani, 2008; Mehra et al., 2001), fewer evidences were found on a positive influence on performance of centrality in the friendship and trust networks. To further explore this association, we hypothesize that:

*H2a: Students who perform better academically are more central in the advice network.*

*H2b: Students who perform better academically are more central in the friendship network.*

*H2c: Students who perform better academically are more central in the trust network.*

## Study Purpose

Although much is known regarding particular associations among the individual variables of the present study (i.e., self-report accuracy, personality traits, centrality and academic performance), investigation of their interrelationships with one another within one comprehensive model is lacking. Figure 1 provides a graphical representation of the variables included in the study as well as the specific hypotheses that we intend to explore.

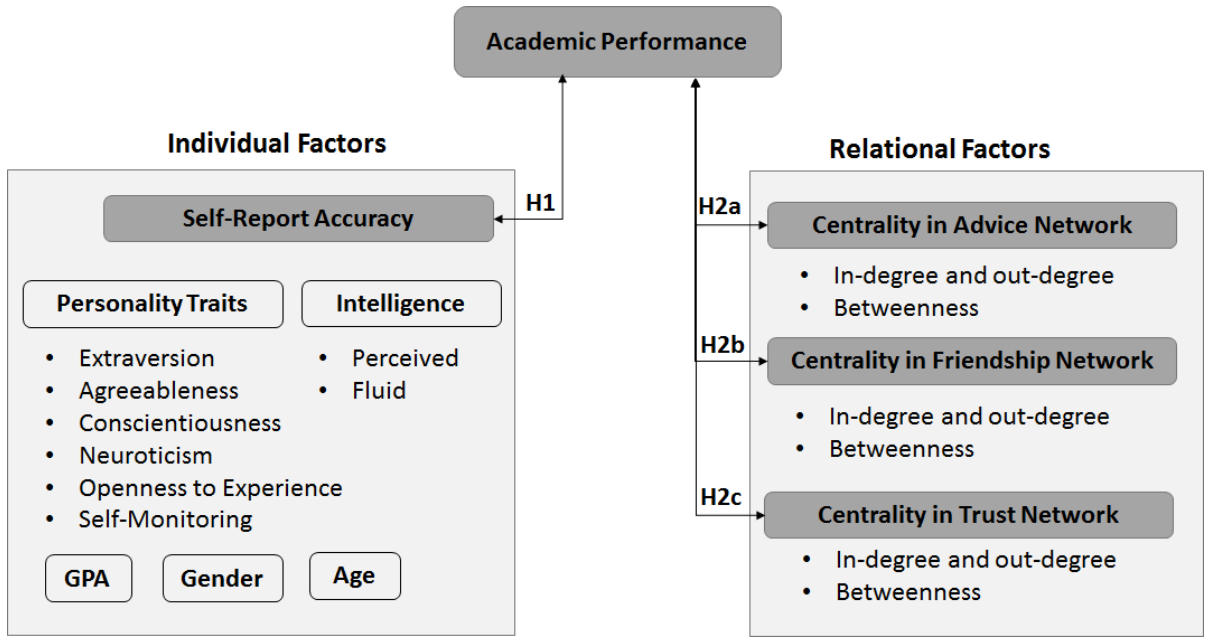

**Figure 1**. Conceptual model.

We included age and current GPA in our model since these variables have been used in previous studies to explain differences in accuracy and cheating behaviors (Franklyn-Stokes & Newstead, 1995; D. L. McCabe et al., 2002; Perry, Kane, Bernesser, & Spicker, 1990) as well as in predicting academic success (Park & Kerr, 1990; Schmitt et al., 2007). For example, previous studies found that male students cheat slightly more often than female students do (Finn & Frone, 2004), though other empirical studies found that gender is not related to cheating or overestimation of academic records (Franklyn-Stokes & Newstead, 1995; Whitley, 1998). For example, Mayer et al. (2007) found an overestimation bias in self-reported SAT scores, where the amount of over-reporting was greater for lower-scoring than higher-scoring students, was greater for upper division than lower division students, and was equivalent for men and women.

## Methodology

### Participants

We carried out a cross-sectional study involving engineering students enrolled in two different courses, a freshmen-year course (Design of Manufacturing Systems) and a senior-year course (Business Management). For both courses students met regularly over four months to attend classes three times per week. The average age of freshman students was 22.62 years ($SD = 1.57$) and 51% of them were male; the average age of senior students was 24.54 years ($SD = 1.62$) and 54% of them were male. Their average GPA was A- for senior students and B- for freshmen students. Every student was extensively informed about our study and expressed willingness to participate, demonstrated by the 100% response rate.

### Instruments and Measures

We used the grade on the final exam of the course to operationalize academic performance, and GPA scores as a proxy for past performance. In total, students took four tests as part of the course and the experiment. Below is a description of the indicators we extracted from these tests.

To assess students' personality traits we used a questionnaire based on the five factor model (Judge et al., 2002) – and we administered a test based on the International Personality Item Pool (Goldberg et al., 2006). In particular, we used the 120-item version of the IPIP-NEO, which covers the traditional five areas. This test has been used extensively and demonstrated to be a reliable measure of the five personality factors (Johnson, Rowatt, & Petrini, 2011; Maples, Guan, Carter, & Miller, 2014). The variable gender has the value of 0 for females and 1 for males.

To compute self-monitoring, we used a version of the Self-Monitoring Scale based on 18-true-false items (Snyder & Gangestad, 1986; Wentzel, 1998). The Cronbach's alpha for this measure was considered reliable (.74).

In order to operationalize intelligence, we adopted Cattell's concept of fluid intelligence (Cattell, 1963) and we used the Raven's test to calculate a score for each student. Specifically, students took one test with the 60-item version of the Raven's Standard Progressive Matrices test, which is a well-known measure of fluid intelligence. This test was used in over 1,000 studies, across different settings and cultures (J. Raven, Raven, & Court, 2000). Past research already proved good retest reliability of this instrument and good internal consistency across different cultural groups (Owen, 1992).

To measure "perceived intelligence" we asked students to express their agreement with the statement "I think this person is intelligent", and rate their peers on a Likert scale – with scores ranging from 0 to 5, where 0 expressed a classification about the statement as "very inaccurate" and 5 as "very accurate". Students could also state they did not know a specific classmate.

In order to map the students' advice, friendship and trust networks, we asked them to indicate respectively: which classmates they usually contacted for advice on course-related topics; which classmates they regularly met outside classrooms (for non-academic purposes); and which classmates they trusted more. Our network surveys, administered in a written form, allowed a free recall, with no limits to the number of names that could be mentioned (Borgatti, Everett, & Johnson, 2013). We built three different network graphs, based on friendship, trust and advice relationships. Each node in the graph is a student of the course and any time a relation is reported, a directed arc is created from a node to another. Arcs have been weighted based on the intensity of the relationship between two nodes. For the advice network, the weight was based on the frequency of the advice seeking interaction and measured on a 5-point Likert scale, where 1 represented interactions happening once or twice during the course, and 5 represented interactions taking place 2 to 3 times a week. For the friendship network, the weight was based on the strength of the friendship relationship and measured using a 3-point scale, where 1 corresponded to friends seen only at the university; 2 was assigned to friends seen also outside of the university; and 3 was assigned to "very close friends". Finally, to build the trust network, we used a 3-point scale where 1 was assigned to people only trusted for university-related matters; 2 was used for people considered trustworthy also on a personal level; and 3 was used for very close people with whom they would share secrets and very intimate details. These three relationships were represented on three separate graphs, where we calculated some of the most commonly used measures of node centrality: degree and betweenness (Everett & Borgatti, 2005; Wasserman & Faust, 1994). The selection of these metrics is consistent with the conceptualization of network centrality as the extent to which a social actor has many direct contacts and is frequently in-between the indirect network paths that interconnect his/her peers.

Degree centrality considers the number of adjacent links to a network node, and expresses the magnitude of direct connections; when weighted, it is calculated as the sum of the weights of the arcs adjacent to a node. To increase its informative power, this measure can be studied considering incoming and outgoing arcs separately. Weighted in-degree centrality sums the values on the arcs terminating at a node; weighted out-degree, on the other hand, sum the values on the arcs that originates from that node and reach the others in the network. Actors with a high in-degree centrality in the advice network are usually preferred for their work or academic-related input, while actors who have a high in-degree centrality in the friendship network are chosen for their companionship and are in general more popular (Fronzetti Colladon et al., 2017; Klein et al., 2004). Students with high values of out-degree centrality in friendship and advice networks ask more often for advices and indicate many more peers as friends. Similarly, students with high in-degree centrality in trust networks are considered trustworthy by their peers; whereas high values of out-degree centrality in this networks indicates people who tend to trust others more, or who built several close relationships over time. Here we focus on students with high values of in-degree and out-degree centrality, which means that they connect with many other students in and out. Another important metric considered in this study is betweenness centrality, which focuses on the capacity of a node to act as an intermediary between any two other nodes. A network is usually highly dependent on actors with high betweenness centrality due to their position as intermediaries and brokers (Everett & Borgatti, 2005). Figure 2 illustrates an example of social network map built on the advice network in both courses. The advice network is sparse and indicates how a relatively high number of students chose to study alone, without asking advices to their peers. This is especially evident in the course with freshmen students.

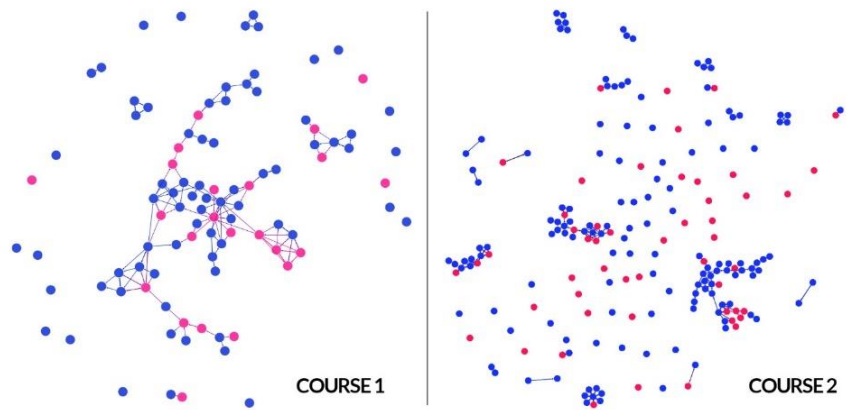

**Figure 2**. Advice Networks - red nodes indicate students reporting more inaccurate data.

**Procedure**

At the beginning of each course we instructed students about the characteristics of our experiment and of the tests they were going to undertake. We made sure to provide detailed instruction to every students, contacting separately those who were absent (2%).

During the first month of the course, students could access any time the computer lab. Proctors helped students complete the online version of the IPIP-NEO questionnaire, saved the result pages, and sent the results to the research team. We chose this approach to preserve the confidentiality of students' responses. Immediately after completing the first test, students were presented a webpage with the 18-item self-monitoring test. Students were not allowed to repeat the tests.

Few days before the end of the course (and the final exam), students participated in the core session of our experiment where they were administered the Raven's Standard Progressive Matrices test containing 60 items (J. Raven et al., 2000). Students took an online version of the test, loaded on a website, and containing 12 items per page. To make sure students' self-report was significantly accurate, they were asked to indicate the time they finished the test. To reduce response bias, we asked them to do so only at the end of the test. This operation was facilitated by the inclusion of a digital real time clock on the webpage. When students reached the final page on the website, a message prompted them to report the time (hours and minutes), with an additional note that "the finishing time would impact the individual test score". To limit cognitive biases in the test results, we included this message only on the final page. Our goal was to create the impression of a possible connection between finishing time and performance, leading to possible deliberate inaccuracies. The website used for the Raven test was automatically recording the actual starting time and finishing time for the test. Any underestimation of the real time was recorded, indicating a possible tendency to be deliberately inaccurate. In this way, we constructed the Inaccuracy variable, expressed in minutes. We also controlled for cases of time overestimation and we found none. The average time automatically reported to complete the Raven's test was 31 minutes. Students were later informed that their test score was never linked to their time inaccuracy. Moreover, no part of the experiment had any influence on their final grades, and students were reassured about this. Lastly, we analyzed the results only after each student completed the final exam of the course (in order not to introduce any evaluation bias on our side). Even if transparently measuring self-report inaccuracy can be difficult, we maintain the importance of taking all the precautions needed to adequately engage with research ethics.

After the Raven's test, students were administered a paper survey with the list of their classmates, with the request to rate their intelligence – as already mentioned, expressing their level of agreement with the statement "I think this person is intelligent". Similarly, they also completed the network surveys presented in the previous section, to report friendship, trust and advice relationships.

To control for extraneous, situational variables related to the environment where the study was conducted, the classrooms where freshmen and senior students took the test had the same spatial characteristics. Standardized instructions were used to ensure that conditions were the same for all participants.

The final exam of the each course was administered on-site, 5 days after the Raven's test, with a two-hour time to complete it for both courses. In the written final exam students had to solve exercises and answer essay questions about course-related topics. Since the two courses were focused on different subjects, the final exams were different. Students' performance was evaluated by the faculty on a common final scale, with grades ranging from 0 to 30 (which is the standard practice in Italian universities).

## Results

The descriptive statistics and t-tests reported in Table 1 indicate interesting trends in terms of who is more likely to be less accurate. Of the 289 students, almost 26% of them were not totally accurate when reporting the finishing time, i.e. their inaccuracy was bigger than 0, with some students underestimating their finishing time up to 40 minutes. Interestingly, there were no cases of overestimation of the finishing time and those who were inaccurate were underestimating of more than 4 minutes. Inaccuracy was not significantly different, on average, for senior students and freshmen. Senior students were on average more extraverted, more cooperative and considerate (high agreeableness) and exhibited higher self-monitoring scores than freshmen. Their grades on the final examination were also higher (B+ vs B-, which translates into the Italian system 26/30 vs 23/30), though this result is not directly comparable since the two courses were covering different topics. All the statistics presented in this paper were calculated using the software Stata.

**Table 1**

*Descriptive Statistics and T-tests*

| Variables | Senior | | Freshmen | | | |
|---|---|---|---|---|---|---|
| Number of Students | 96 | | 193 | | | |
| % Male | 54.17% | | 51.30% | | | |
| Variables | Mean | SD | Mean | SD | t | p |
| Inaccuracy | 1.45 | 2.78 | 2.05 | 4.50 | 1.09 | .276 |
| Age | 24.54 | 1.62 | 22.62 | 1.57 | 9.71 | .000 |
| Self-Monitoring | 8.45 | 2.79 | 7.64 | 3.19 | 2.12 | .035 |
| Extraversion | 57.19 | 17.99 | 52.15 | 16.33 | 2.39 | .018 |
| Agreeableness | 41.09 | 20.12 | 34.71 | 15.68 | 2.96 | .003 |
| Conscientiousness | 52.26 | 17.23 | 50.18 | 18.00 | .94 | .348 |
| Neuroticism | 37.63 | 17.41 | 41.12 | 15.61 | -1.72 | .086 |
| Openness to Experience | 39.10 | 17.83 | 40.34 | 16.88 | -.58 | .565 |
| Past Performance (GPA) | 26.74 | 2.14 | 23.99 | 1.48 | 12.80 | .000 |
| Exam Grade | 26.04 | 3.66 | 23.94 | 2.90 | 5.30 | .000 |
| Fluid Intelligence | 51.64 | 6.72 | 52.54 | 4.73 | -1.33 | .184 |
| Perceived Intelligence | 3.59 | .60 | 3.44 | .65 | 1.85 | .065 |

Note. T-test are used to test the significance of the differences between senior students and freshmen.

Table 2 shows the correlations among variables. Given the high number of correlations, we want to be conservative and consider an alpha level at least lower than .01 to indicate significance. The table indicates a significant negative association between inaccuracy and performance at the final exam. Exam performance is also correlated negatively with openness to experience and positively with extraversion, perceived intelligence and betweenness centrality in the advice network. Past performance, is significantly and positively correlated with exam performance, conscientiousness and perceived intelligence; by contrast, it negatively associates with neuroticism. In addition, past performance is also significantly correlated with students' betweenness centrality in all networks, and with the weight of incoming and outgoing ties in advice networks (representing the tendency of students to ask and being asked for advices on course-related topics). We also noticed a positive correlation between inaccuracy and extraversion, self-monitoring and network centrality metrics (for friendship and trust relationships). Extraversion is correlated with higher centrality in friendship networks, both in terms of betweenness centrality and in terms of incoming ties (number of friends reporting being friends with ego). In addition, extraversion positively correlates with betweenness centrality in trust relationships. Centrality measures in all the networks were strongly correlated.

**Table 2**

*Pearson correlation coefficients*

| Variables | | 1 | 2 | 3 | 4 | 5 | 6 | 7 | 8 | 9 | 10 | 11 | 12 | 13 | 14 | 15 | 16 | 17 | 18 | 19 | 20 | 21 | 22 |
|---|---|---|---|---|---|---|---|---|---|---|---|---|---|---|---|---|---|---|---|---|---|---|---|
| 1 | Inaccuracy | 1.00 | | | | | | | | | | | | | | | | | | | | | |
| 2 | Gender | .02 | 1.00 | | | | | | | | | | | | | | | | | | | | |
| 3 | Age | -.02 | .00 | 1.00 | | | | | | | | | | | | | | | | | | | |
| 4 | Self-Monitoring | .17** | .25** | .13* | 1.00 | | | | | | | | | | | | | | | | | | |
| 5 | Extraversion | .16** | .01 | .06 | .42** | 1.00 | | | | | | | | | | | | | | | | | |
| 6 | Agreeableness | .08 | .25** | .09 | -.12* | .12* | 1.00 | | | | | | | | | | | | | | | | |
| 7 | Conscientiousness | .02 | .19** | -.17** | .01 | .08 | .26** | 1.00 | | | | | | | | | | | | | | | |
| 8 | Neuroticism | .03 | -.02 | -.03 | -.10 | -.32** | -.07 | -.41** | 1.00 | | | | | | | | | | | | | | |
| 9 | Openness to Experience | .16** | -.01 | .00 | .14* | .29** | .17** | .04 | -.05 | 1.00 | | | | | | | | | | | | | |
| 10 | Past Performance | -.10 | -.09 | .21** | .07 | .10 | .07 | .20** | -.15** | -.14* | 1.00 | | | | | | | | | | | | |
| 11 | Exam Performance | -.45** | -.01 | .09 | .12* | .17** | -.06 | .09 | -.15* | -.16** | .53** | 1.00 | | | | | | | | | | | |
| 12 | Fluid Intelligence | .05 | .04 | -.28** | .00 | -.04 | .15** | -.03 | .15* | -.01 | .04 | .04 | 1.00 | | | | | | | | | | |
| 13 | Perceived Intelligence | -.05 | .07 | -.07 | -.14* | -.02 | .07 | .15* | -.24** | -.20** | .35** | .23** | .27** | 1.00 | | | | | | | | | |
| 14 | Weighted In-degree (Advice) | .15* | -.05 | -.07 | .04 | .06 | -.01 | .11 | -.08 | -.14* | .34** | .14* | .21** | .46** | 1.00 | | | | | | | | |
| 15 | Weighted Out-degree (Advice) | .10 | -.03 | .13* | .10 | .09 | -.01 | -.09 | .02 | -.03 | .17** | .12* | -.01 | .14* | .29** | 1.00 | | | | | | | |
| 16 | Betweenness Centrality (Advice) | .12* | -.08 | .05 | .01 | .10 | .04 | .12* | -.06 | -.13* | .29** | .20** | .13* | .21** | .51** | .53** | 1.00 | | | | | | |
| 17 | Weighted In-degree (Friendship) | .16** | -.04 | -.09 | .12* | .22** | -.07 | .06 | -.15* | -.03 | .15* | .05 | .12* | .23** | .48** | .37** | .39** | 1.00 | | | | | |
| 18 | Weighted Out-degree (Friendship) | .14* | -.05 | -.14* | .09 | .14* | -.02 | .08 | -.05 | .04 | .04 | -.02 | .09 | .18** | .37** | .42** | .31** | .75** | 1.00 | | | | |
| 19 | Betweenness Centrality (Friendship) | .19** | -.04 | -.01 | .14* | .18** | .02 | .15* | -.10 | .04 | .23** | .09 | -.07 | .08 | .40** | .41** | .44** | .66** | .73** | 1.00 | | | |
| 20 | Weighted In-degree (Trust) | .17** | -.10 | -.22** | .02 | .10 | -.11 | .04 | -.13* | -.02 | .03 | -.06 | .17** | .32** | .48** | .27** | .31** | .78** | .68** | .44** | 1.00 | | |
| 21 | Weighted Out-degree (Trust) | .16** | -.04 | -.17** | .06 | .11 | -.03 | .05 | -.06 | .07 | -.04 | -.04 | .14* | .20** | .35** | .40** | .24** | .65** | .79** | .54** | .72** | 1.00 | |
| 22 | Betweenness Centrality (Trust) | .16** | .02 | .08 | .07 | .18** | .09 | .14* | -.09 | -.07 | .25** | .08 | .01 | .22** | .40** | .36** | .44** | .65** | .60** | .65** | .59** | .59** | 1.00 |

*p < .05. **p < .01. ***p < .001.

Being accurate and reporting the correct finishing time is strongly associated with student performance, at least with regards to the final examination: those who honestly reported the finishing time as it appeared on the screen (zero inaccuracy), exhibited a significantly higher performance ($M = 25.65$, $SD = .18$) than those who were less accurate ($M = 21.72$, $SD = .41$), $t(287) = 10.25$, $p = .000$. In addition, betweenness centrality seems to be associated with higher final grades, in the advice network.

Students seemed to connect with others exhibiting similar attributes in terms of age and gender – a phenomenon known as network homophily (Mcpherson, Smith-Lovin, & Cook, 2001). A clustering tendency is not directly quantifiable from the social network maps (Figure 2). A better measure to indicate the magnitude of network correlation with respect to students' personal attributes is offered by the Geary's coefficients reported in Table 3 (Geary, 1954). Values equal to 1 indicate perfect independence, while values close to 0 indicate strong attribute homogeneity (or homophily). Consequently, values bigger than 1 indicate a tendency for social actors to connect with others with different attribute scores. The significance of coefficients was tested considering distributions generated from 30,000 random permutations and by using a quadratic assignment procedure (QAP) (Krackhardt, 1988). This procedure is necessary here, as network matrices are being correlated, facing a problem of non-independence of network ties. For example, if actor A is friend with actor B and C, he could introduce B to C and support the creation of a friendship tie between them. In such a scenario, the existence of a friendship relationship between B and C (i.e. a tie) could be more probable than a link between B and a generic node D unconnected with the others. The QAP procedure solves the problem of non-independence of network observations: the evaluation of the significance of the correlation coefficients is made considering a non-parametric distribution which originates from the random permutations of rows and columns in the matrices (Krackhardt, 1988).

Table 3 indicates that students who are less accurate in self-reporting data do not rely on each other in terms of advice seeking and in terms of friendship. We also observe a tendency for students to get together with similar others based on gender, age and GPA.

**Table 3**

*Students' homophily (Geary's C)*

| Variables | Senior Students | | | Freshmen Students | | |
|---|---|---|---|---|---|---|
| | Advice | Friendship | Trust | Advice | Friendship | Trust |
| Gender | .74*** | .68*** | .66*** | .68*** | .72*** | .51*** |
| Age | .78* | .87 | .96 | .16*** | .19*** | .17*** |
| Self-Monitoring | .74** | .88 | .73** | 1.12 | 1.11* | 1.05 |
| Extraversion | 1.24* | 1.12 | 1.30** | 1.04 | 1.12 | 1.09 |
| Agreeableness | 1.08 | 1.10 | 1.20* | .98 | .98 | .97 |
| Conscientiousness | .76* | .94 | .89 | 1.16 | 1.01 | 1.01 |
| Neuroticism | 1.43*** | 1.10 | 1.19* | 1.12 | 1.16* | 1.16* |
| Openness to Experience | 1.14 | 1.09 | 1.19 | 1.08 | 1.27*** | 1.29*** |
| Past Performance | .69** | .75** | .68** | 1.10 | .79*** | .70*** |
| Exam Performance | 1.42** | 1.06 | 1.15 | 1.08 | 1.24** | 1.11 |
| Inaccuracy | 1.48** | 1.27** | 1.32** | 1.63* | 1.40* | 1.40* |
| Fluid Intelligence | 1.04 | .99 | .72 | .78 | 1.13 | 1.10 |
| Perceived Intelligence | .85 | .86 | .85 | .67*** | .69*** | .64*** |

*p < .05. **p < .01. ***p < .001.

Table 4 shows the hierarchical multiple regression models we used to identify the main determinants of performance.

**Table 4**

*Predicting performance: multiple regression models*

| Variables | Model 1 | Model 2 | Model 3 | Effect Size (Model 3) $\omega^2$ | CI 95% |
|---|---|---|---|---|---|
| Intercept | 3.723 | 3.359 | 4.097 | | |
| Gender | .469 | .357 | .256 | 0 | 0-.024 |
| Age | -.005 | -.040 | -.024 | 0 | 0-.012 |
| Course 2 | -.188 | -.446 | -.549 | .001 | 0-.031 |
| Past Performance | .805*** | .834*** | .714*** | .168 | .093-.248 |
| Self-Monitoring | -.001 | .015 | .083 | .004 | 0-.038 |
| Extraversion | .031* | .032** | .038*** | .044 | .007-.103 |
| Agreeableness | -.021 | -.024* | -.016 | .007 | 0-.044 |
| Conscientiousness | -.005 | -.005 | -.002 | 0 | 0-.012 |
| Neuroticism | -.009 | -.010 | -.003 | 0 | 0-.013 |
| Openness to Experience | -.022* | -.021* | -.013 | .004 | 0-.037 |
| Fluid Intelligence | .021 | .021 | .030 | 0 | 0-.028 |
| Perceived Intelligence | .084 | .480 | .378 | .002 | 0-.034 |
| Weighted Indegree (Advice) | | -.912 | -.386 | 0 | 0-.022 |
| Weighted Outdegree (Advice) | | .076 | .211 | 0 | 0-.016 |
| Betweenness Centrality (Advice) | | 1.098 | 1.148* | .011 | 0-.052 |
| Weighted Indegree (Friendship) | | .755 | .546 | 0 | 0-.019 |
| Weighted Outdegree (Friendship) | | -.745 | -1.594 | .005 | 0-.042 |
| Betweenness Centrality (Friendship) | | -.074 | .753 | 0 | 0-.027 |
| Weighted Indegree (Trust) | | -2.598* | -1.877 | .010 | 0-.050 |
| Weighted Outdegree (Trust) | | 1.667 | 1.624 | .010 | 0-.049 |
| Betweenness Centrality (Trust) | | -.459 | -.230 | 0 | 0-.015 |
| Inaccuracy | | | -.331*** | .256 | .171-.338 |
| Adjusted R-squared | .302 | .320 | .494 | | |
| R-Squared | .331 | .370 | .533 | | |

| | Change in R-Squared | | |
|---|---|---|---|
| | Difference | F | p |
| Model 2 - Model 1 | .039 | 1.822 | .064 |
| Model 3 - Model 2 | .163 | 92.992 | .000 |

*Note*. N = 289. *p < .05. **p < .01. ***p < .001.

In Model 1, we included all predictors, except centrality measures and inaccuracy. This model could explain about 33% of variance. Adding centrality measures in Model 2 led to an increase of the $R^2$ of about 4%. Finally, the inclusion of the inaccuracy variable in Model 3 led to an increase of $R^2$ of about 16%, with 49% of variance explained overall.

Table 4 also reports the effect size of each predictor in Model 3, using the omega-squared estimator (Albers & Lakens, 2017). These results show that inaccuracy had the biggest effect ($\omega^2 = .26$) followed by past performance ($\omega^2 = .17$), extraversion ($\omega^2 = .04$) and betweenness centrality in the advice network ($\omega^2 = .01$), with the latter having an almost negligible effect. All the other predictors are not significant and have a close-to-zero effect size. The major predictors of academic performance are therefore self-report accuracy and past performance (GPA). It seems that accurate students with a better past academic performance also earn higher final exam grades. In addition, highly performing students seem to be more extraverted and more central in the advice network, which gives them the opportunity to acquire knowledge from multiple sources. On the other hand, neither perceived intelligence nor fluid intelligence were significantly influencing performance. Similarly, centrality in trust and friendship networks does not show a significant influence on performance. Lastly, we used a dummy variable to control for the course in which students were enrolled, as different course topics and different final exams might affect performance. However, this variable was also not significant.

## Discussion

The results of our exploratory study indicate that higher achieving students tend to be more accurate when asked real time to report data that might or might not impact how they are perceived by instructors. Figure 3 summarizes the results of the multiple regression models and highlights the variables that are more likely to predict performance.

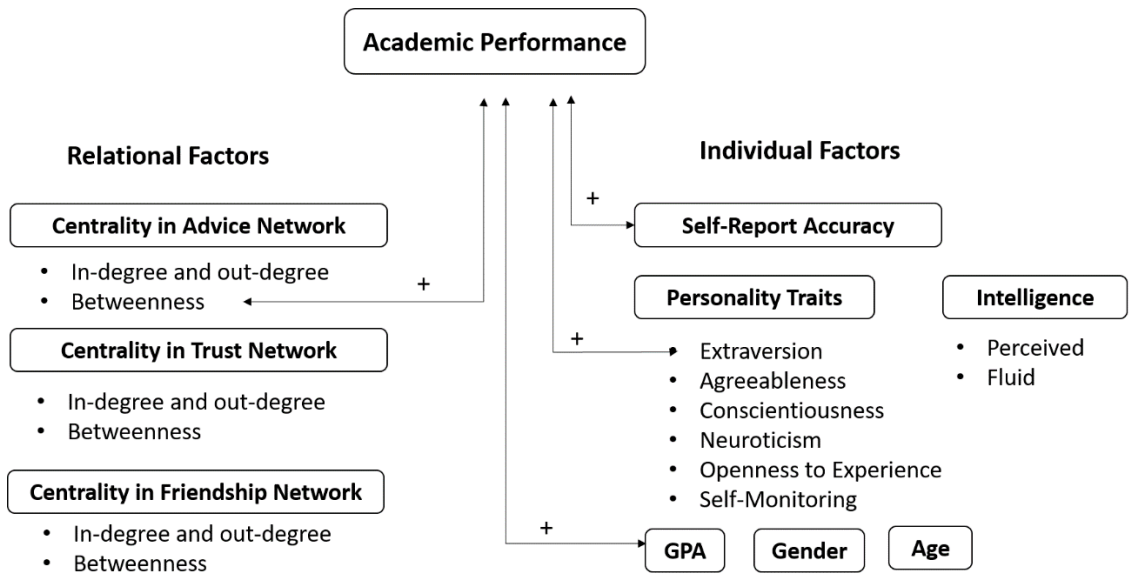

**Figure 3**. Predictors of academic performance.

It seems that students who are less confident in their academic ability or have difficulties with the subject of the course might have reported a shorter time to finish the test in order to look better, and to project a positive image (Tracey, 2016). This tendency to "look better" and to be concerned with self-presentation is a typical characteristic of people with high self-monitoring traits. In our study, we found confirmation to this with a positive correlation between inaccuracy and self-monitoring: individuals with a high self-monitoring personality trait tend to regulate their behavior to accommodate social situations, which could explain the need to report an inaccurate finishing time to the instructor (Beretvas, Meyers, & Leite, 2002; Mehra et al., 2001).

We also found a positive correlation of extraversion with inaccuracy. Extraverted students probably were more driven by the desire to spend time with others after class, and they were less inclined to double check the exact time. A possible motivation of this lack of accuracy could be their need to join the rest of the class after a very long examination. Contrary to previous studies (Busato et al., 2000; Conard, 2006), only extraversion (and not neuroticism, openness to experience, agreeableness and conscientiousness) seems to be associated with academic performance. Extraverted students usually play the intermediary role of information brokers and have more ties to others, which might give them an advantage in gathering important information to succeed in the final exam.

Consistent with previous research (Mehra et al., 2001; Tasselli & Kilduff, 2017), our study found that centrality measures could be used to predict performance, even though the models indicate a smaller influence of centrality than expected. Results seem to suggest that performance is not higher for individuals who are central in the friendship network, though it is for students who act as brokers in the advice network by showing higher betweenness centrality. Although the effect size of betweenness centrality is small, the positive association between higher final grades and betweenness centrality in the advice network could be explained by considering the increased ability for student-brokers to access different knowledge sources and recombine this knowledge to increase their understanding of the course topics. Students who are more central in the advice network might earn better grades as they interact with several other peers exchanging information and offering their advice. This is consistent with recent studies showing the role of advice-seeking networks as valuable resource for students (Battistoni & Fronzetti Colladon, 2014).

In addition, while being more central and more connected to peers seeking and offering advices could lead to higher grades on the final exam, being well-connected in the trust and friendship networks does not predict your chances of earning a good grade. This contrasts with other empirical studies indicating the role of peer support and friendship in increasing academic performance and motivation (Altermatt & Pomerantz, 2003; Buhs, 2005; Furrer & Skinner, 2003). At the same time, previous studies do not unequivocally support the association between centrality in the friendship network and academic success. As Baldwin noted (Baldwin et al., 1997) there is no clear evidence in literature that stronger friendship ties with peers lead to higher academic accomplishments. Having more friends you can trust might not be enough to help you get all the information you need to excel at the exam. Relying on the emotional support of friends might help on a psychological level, but only those ties to people who have specialized knowledge can have an impact on performance.

Contrary to previous studies, we found no evidence that intelligence – either perceived or fluid - could predict the grade on the final examination (Busato et al., 2000; Furnham, Zhang, & Chamorro-Premuzic, 2006). A possible reason could be the way we operationalized intelligence, which differs from other studies. Intelligence has been traditionally measured using student's percentile ranking on the ACT or the student's cumulative GPA (Park & Kerr, 1990), or using a combination of self-assessment, Baddeley Reasoning Test and Wonderlic Personnel Test (Furnham et al., 2006). In this study, we only used the Raven's test to measure fluid intelligence, which could be a possible reason for the results. Building a more inclusive metric of intelligence, integrating some of the personality traits we observed, could have been a better option and should be considered when replicating this study.

Lastly, controlling for age and gender did not modify our findings, which is aligned with previous studies (Finn & Frone, 2004).

## Conclusions and Limitations

The goal of this study was to explore possible associations between academic performance and multiple individual and relational variables. The results contribute to the literature on self-report accuracy by exploring a new measure, precision at reporting the finishing time, which has not been traditionally collected and correlated to performance. Differently from previous studies (Cole & Gonyea, 2010), we do not measure accuracy looking at self-reported academic records (e.g. GPA, GRE, STA, ACT). Instead, we use the accuracy in reporting the finishing time on a cognitive test, which might not necessarily trigger a social desirability effect.

Asking students to self-report academic records in higher education is a widespread method to collect performance data and its validity has been widely debated for more than forty years (Pike, 2011). In this study, we suggest to use a new indicator, self-report accuracy on a test finishing time, which seems to be highly correlated with academic performance. The results are aligned with existing literature emphasizing the individual and relational factors associated with academic performance, and pending future studies, may be utilized to include new metrics of self-report accuracy. Our study also included several centrality metrics across three different networks, instead of focusing on one single network (e.g. advice or trust). For example, Cattani and Ferriani (2008) looked at variables such structural holes and coreness to predict creative outcomes, and did not include centrality measures like betweenness and degree centrality, which can offer additional insights on the impact of relational ties on performance.

Instead of looking at the impact of a single social network on behavior and performance, as previous studies have done (Mehra et al., 2001; Tasselli & Kilduff, 2017), we prefer to include all the possible connections among peers that can support them emotionally (trust and friendship networks) and professionally (advice network). We believe that combining the three networks of friendship, advice and trust we can better depict their reciprocal influence and better understand the impact on individual behavior and performance.

By offering a comprehensive look at centrality, accuracy and performance, this study uncovers correlations between variables impacting academic accomplishments that were never considered together. While previous studies have explored associations between accuracy at recalling people and social network roles (Grippa & Gloor, 2009), fewer studies have explored the association between accuracy at reporting an academically-related information (exam finishing time) and centrality in the friendship, trust and advice networks.

The specific contribution of this study is twofold. First, we demonstrated that a higher self-report accuracy is correlated with academic success, and we do this by looking at the discrepancy between a digitally recorded finishing time and a factual finishing time (self-reported data). Second, we explored the associations between multiple indicators of social network centrality within three different networks and found that being central in the advice network is correlated with higher academic performance, but with a small effect in regression models. This study looks at centrality indicators that are not always associated with performance, behaviors and traits.

For researchers and education administrators seeking to use student self-reported data in their work, a way to avoid the overestimation bias is to reflect on other instruments that are not directly related to student academic records. To reduce the impact of misrepresentation and encourage more accuracy, researchers need to include methods of control such as prompts in the instrument, indicating that responses will be cross-checked with actual data (e.g. transcripts), similarly to what we have done in this study.

This study offers empirical evidences to help educators, administrators and education consultants recognize the multiple variables that influence academic achievement. While individual characteristics have been demonstrated for decades to impact performance (e.g. gender, personality traits, parents' level of education, presence of positive role models, valuing education), a relational-interpersonal dimension should be considered and prompt action in the classroom. Improvement actions for educators include facilitating study groups, increase the number of team assignments, encourage outside of classroom activities, and incorporate team building activities in the curriculum.

A limitation of this study is related to the sample of engineering students that we involved, all affiliated with a single higher-education institution. This limited our ability to draw significant conclusions about contextual influences and about the role of specific disciplines (D. L. McCabe et al., 2002). We suggest replicating our study on a larger sample, in different institutions and possibly investigating the impact of cultural and national differences.

It is important to recognize that other predictors may also play an important role in determining performance, which require further investigation. These factors include connections outside of the classroom environment, such as membership to student associations or study groups. Since the sample included students who had attended other courses together, future research should differentiate between newly established and old ties. Finally, one should not underestimate the impact of test anxiety on the accuracy at reporting data, especially since we measured accuracy after students took an intelligence test. In a study involving more than 5000 US graduate and undergraduate students, Chapell and colleagues (2005) found that test anxiety is associated with reductions in GPA both at the undergraduate and graduate level. In particular, lower-test-anxious female students earned a higher GPA. Since academic performance is influenced by a multiplicity of factors, future studies should also include metrics like the TAI - Test Anxiety Inventory (Spielberger, 1987), which could help understand whether self- report accuracy was influenced by other factors.

Uppal, N. (2014). Moderation effects of personality and organizational support on the relationship between prior job experience and academic performance of management students. *Studies in Higher Education*, *39*(6), 1022–1038. https://doi.org/10.1080/03075079.2013.777411

Warris, A., & Rafique, R. (2009). Trust in Friendship: A Comparative Analysis of Male and Female University Students. *Bulletin of Education & Research*, *31*(2), 75–84.

Wasserman, S., & Faust, K. (1994). *Social Network Analysis: Methods and Applications*. New York, NY: Cambridge University Press. https://doi.org/10.1525/ae.1997.24.1.219

Wentzel, K. R. (1998). Social relationships and motivation in middle school: The role of parents, teachers, and peers. *Journal of Educational Psychology*, *90*(2), 202–209. https://doi.org/10.1037/0022-0663.90.2.202

Wentzel, K. R. (2005). Peer relationships, motivation, and academic performance at school. In A. J. Elliot & C. S. Dweck (Eds.), *Handbook of competence and motivation* (pp. 279–296). New York, NY: Guilford Publications.

Whitley, B. E. (1998). Factors associated with cheating among college students: A review. *Research in Higher Education*, *39*(3), 235–274. https://doi.org/10.2307/40196379

Willard, G., & Gramzow, R. H. (2008). Exaggeration in memory: Systematic distortion of self-evaluative information under reduced accessibility. *Journal of Experimental Social Psychology*, *44*(2), 246–259. https://doi.org/10.1016/j.jesp.2007.04.012